\documentclass[12pt]{article}
\usepackage{palatino}
\usepackage{mathpazo} 

\usepackage{pyz}
\usepackage{booktabs} 
\usepackage[ruled]{algorithm2e} 
\usepackage{clipboard}
\usepackage{tikz}
\usepackage{url}
\usepackage{natbib}
\usepackage{amsthm,amsmath,amssymb}
\usepackage{comment}
\usepackage{booktabs}

\SetAlFnt{\small}
\SetAlCapFnt{\small}
\SetAlCapNameFnt{\small}
\SetAlCapHSkip{0pt}
\IncMargin{-\parindent}

\newcommand*{\surjective}{surjective}

\newtheorem{assumption}{Assumption}
\newtheorem{proposition}{Proposition}[section]
\newtheorem{observation}{Observation}[section]

\setcitestyle{authoryear}

\usepackage{setspace}
\begin{document}
\title{Distributionally Robust Principal Agent Problems and Optimality of Contracts}

\author{Peter Zhang \\ Heinz College, Carnegie Mellon University \\ pyzhang@cmu.edu \\ \emph{\small{First Version: March 2023}} \\ \emph{\small{Current Version: January 2024}}}
\date{}

\maketitle

\begin{abstract}
We propose a distributionally robust principal agent formulation, which generalizes some common variants of worst-case and Bayesian principal agent problems. We construct a theoretical framework to certify whether any \surjective{} contract family is optimal, and bound its sub-optimality. We then apply the framework to study the optimality of affine contracts. We show with geometric intuition that these simple contract families are optimal when the surplus function is convex and there exists a technology type that is simultaneously least productive and least efficient. We also provide succinct expressions to quantify the optimality gap of any surplus function, based on its concave biconjugate. This new framework complements the current literature in two ways: invention of a new toolset; understanding affine contracts' performance in a larger landscape. Our results also shed light on the technical roots of this question: why are there more positive results in the recent literature that show simple contracts' optimality in robust settings rather than stochastic settings? This phenomenon is related to two technical facts: the sum of quasi-concave functions is not quasi-concave, and the maximization and expectation operators do not commute. 
\end{abstract}

\section{Introduction}\label{sec:intro}

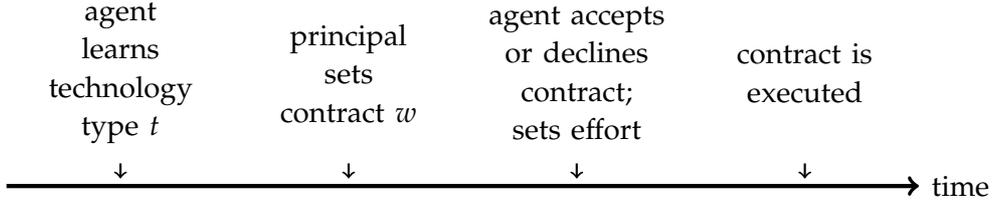
\begin{figure}[ht]
\centering
\begin{tikzpicture}
\draw[->,ultra thick] (-6,0)--(6,0) node[right]{\small{time}};

\draw[->, thick] (-4.5,0.3)--(-4.5,0.1);
\draw[->, thick] (-1.5,0.3)--(-1.5,0.1);
\draw[->, thick] (1.5,0.3)--(1.5,0.1);
\draw[->, thick] (4.5,0.3)--(4.5,0.1);

\draw(-4.5,1.5) node[text width=2cm,align=center] {\small{agent learns technology type $t$}};
\draw(-1.5,1.5) node[text width=2cm,align=center] {\small{principal sets contract $w$}};
\draw(1.5,1.5) node[text width=2.5cm,align=center] {\small{agent accepts or declines contract; sets effort}};
\draw(4.5,1.5) node[text width=2cm,align=center] {\small{contract is executed}};
\end{tikzpicture}
\caption{Sequence of the game.}\label{fig:sequence}
\end{figure}

A principal agent problem is a Stackelberg game involving a principal (she) and an agent (he). The principal makes the first move and sets a contract. Then, if the agent accepts the contract, he chooses an effort level that produces some output. The principal's payoff is the difference between this output and the payment sent to the agent, and the agent's payoff is the payment received from the principal minus the cost of the chosen effort level (Figure \ref{fig:sequence}).

We describe the principal's and agent's decision problems as follows.

\begin{equation} \label{eq:DRO-principal}
    \text{Principal's Problem:} \quad \max_{w \in \mathcal{W}} \min_{G \in \mathcal{G}} \mathbb{E}_{t \sim G} \left[ \mathbb{E}_{y \sim F_{t,w}} [u(y) - w(y)] \right],
\end{equation}

\begin{equation} \label{eq:DRO-agent}
    \text{Agent's Problem:} \quad \max_{(F,c) \in \mathcal{A}_t} \mathbb{E}_{y \sim F}[w(y)] - c.
\end{equation}

The principal takes a \emph{distributionally robust} perspective, maximizing her outcome under the worst technology type distribution. 
The principal's main decision is the contract $w$, which is a function that maps observed output to payment. $\mathcal{W}$ is the contract family she chooses from. 
We call the minimization and expectation ``$\min_{G \in \mathcal{G}} \mathbb{E}_{t \sim G}$'' together as the \emph{worst-expectation} operator. The worst distribution is taken over all potential technology type distributions: $t$ is the technology type, $G$ is some technology type probability distribution, and $\mathcal{G}$ is an ambiguity set that contains all plausible type distributions. Within the principal's objective function: $y$ is the realized output level; $F_{t,w}$ is from the maximizer of the Agent's Problem (\ref{eq:DRO-agent}), representing the random distribution of output; $u$ is the principal's utility function.

\paragraph{Principal's Distributional Robust Perspective} Note that we make little assumption about the structure of $\mathcal{G}$. For example, if $\mathcal{G}$ is a singleton, then the problem reduces to a Bayesian setting where the principal cares about the expected outcome. If $\mathcal{G}$ is a set that contains all distributions including the delta distributions, then this problem reduces to a worst-case robust setup. In recent literature, \citet{Carroll2015}, \citet{dutting2019}, and \citet{Yu2020} (implicitly) adopt this distributionally robust paradigm, with specific structures on $\mathcal{G}$. While imposing structure on $\mathcal{G}$ is by far the most common approach and helps with theoretical analysis (in both these three contract design papers and the entire optimization literature), we take an different approach to not impose structure on $\mathcal{G}$. We show that somewhat surprisingly, we can still derive useful results, which complement the existing literature in interesting ways.

\paragraph{Agent's Decision and Technology Type} For each technology type $t$, we assume the agent has a decision space $\mathcal{A}_t \ni (F,c)$, where each $c$ is the cost of an effort level and $F$ is a probability distribution of the output. We call $\mathcal{A}_t$ the type $t$ \emph{technology} \citep{Carroll2015}. When a type $t$ agent (agent with technology type $t$) accepts contract $w$, we denote his optimal decision as $(F_{t,w}, c_{t,w})$. We note that while our terminology ``type'' is mathematically consistent with what the literature does when modeling information asymmetry (here the principal does not know the agent's technology set), we do not go into the usual discussion of screening, revelation principle, or contract \emph{menus}. Nonethess, it can be seen that our framework and optimality results provide a conservative upper bound for the case of principal providing contract menus, since a game with screening is sandwiched between Games I and III shown in our framework below. 


\paragraph{On Risk Aversion} The contract theory literature usually incorporates some type of risk aversion, either via a risk averse agent utility function or via a limited liability assumption where there is a lowerbound (\eg, zero) on how much the principal should pay the agent under all situations. We take the second approach.

\subsection{Relationship to the Literature}\label{subsec:lit-contributions}

We first focus on some recent papers that are most related to this current work. We then provide a general and brief overview of the principal agent literature.

The current work is most related to four recent papers: \citet{Carroll2015, dutting2019, Yu2020, Li2022}. 
\citet{Carroll2015} shows the optimality of affine contracts in a robust setting: the principal knows only a subset of actions available to the agent, and wants to find worst-case optimal contracts against such uncertainty. The author finds that affine contracts are worst-case optimal. 
\citet{dutting2019} study the (near) optimality of linear contracts. The authors show that linear contracts are optimal when the principal knows the first moment of each effort level's output distribution. They also show tight approximation bounds for linear contracts from a worst-case approximation perspective. 
In the setting of \citet{Yu2020}, the principal knows the support of the effort-contingent output probability mass distribution. The principal's problem is to choose a contract on the output level in order to maximize her worst-case payoff among all potential parameter uncertainties, subject to the contract being worst-case incentive compatible. The authors show that affine contracts are optimal. These three aforementioned papers all incorporate distributional robustness in some form, with specific structures on uncertainty. In contrast, we do not impose any structure, and derive complementary results.
\citet{Li2022} study an instance of the principal agent model in a forest conservation setting, where the principal, being a policymaker, is uncertain about the agent's reservation effort level and wants to design a policy to maximize expected conservation utility. The authors show that affine contracts are guaranteed to achieve at least 50\% of the maximum second-best utility. 

In terms of analysis, this paper brings a theoretical perspective that has not appeared in the contract design literature, and utilize modern tools that are emerging from the dynamic optimization literature.  
In particular, this study joins a growing literature in the analysis of the so-called decision rules in dynamic robust optimization.
\citet{ben2004adjustable} proposes affine decision rules to tractably and approximately solve dynamic robust optimization problems. \citet{Iancu2013} provides the most general characterization of the optimality of affine decision rules to date. Other theoretical results for the (near) optimality of affine decision rules have been proven by \citet{chen2008linear}, \citet{bertsimas2011geometric},
\citet{ardestani-jaafari2016},
\citet{sns},
\citet{el2021optimality},
\citet{WeiZhang2022}, and
\citet{LuSturt2022}.
Decision rules in the dynamic robust optimization problem are equivalent to contracts in the principal agent problem.
A key difference between the two problems is that dynamic robust optimization is about sequential zero-sum games, whereas principal agent models are sequential general-sum games.


Contract theory and principal agent models are historically studied in the economics literature. A fundamental task is to align incentives between the principal and the agent under information asymmetry \citep{holmstrom1979moral}. For example, how should a principal design contracts to screen and select agents with the right skills and attributes (related to adverse selection) \citep{mirrlees1976optimal}? How should the principal design contracts to mitigate moral hazard, where the agent could potentially take unobservable actions that are not in the best interest of the principal \citep{grossman1983analysis}? Other issues include risk sharing \citep{innes1990limited}, the impact of long-term relationships on contract design \citep{spear1987repeated}, multitasking \citep{holmstrom1991multitask}, multiple agents \citep{demski1984}, and so on. For a comprehensive treatment on the subject, we refer the readers to \citet{Laffont2009}. Related to our applications section, there has been a growing discussion about simple contracts, in many cases in the robust setting \citep{Holmstrom1987, Edmans2011, Chassang2013, Garrett2014, Carroll2015, Carroll2016, Tang2021, castiglioni2022linear, WaltonCarroll2022}.

In the operations management literature, information asymmetry in a supply chain setting comes from one party having more information than the other about demand, cost, quality, effort, or lead time \citep{lariviere1999supply, tsay1999quantity, dana2001revenue, cachon2003supply, tomlin2003, corbett2004designing, chen2007, perakis2007, kayaOzer2009, zhangfuqiang2010, kayis2013, kimNetessine2013}. Similar to the economics litereature, there is also a growing stream of research to understand simple contracts \citep{Chen2005, CachonZhang2006, zhangfuqiang2010, kayis2013,DuenyasHuBeil2013, ChenLaiXiao2016, Yu2020, Li2022}. 

In the algorithmic game theory literatuer, principal agent problems have seen growing interests. Topics include the complexity of optimal contracts, or conversely, the optimality of simple contracts \citep{dutting2019, dutting2021, guruganesh2021, castiglioni2022random}; the benefit of randomization \citep{castiglioni2022random}; and the sample complexity of contract design in a dynamic setting \citep{zhu2023}.

\subsection{Contributions}

The paper introduces a distributionally robust principal agent formulation, expanding upon existing models in principal agent problems by incorporating both worst-case and Bayesian perspectives. This approach allows for minimal assumptions about the structure of technology type distribution uncertainty, offering a more general framework compared to previous studies. Key Contributions include:

\begin{enumerate}
    \item Theoretical Framework for Contract Optimality: This paper establishes a theoretical framework to determine the optimality of any surjective contract family and to bound its sub-optimality. This framework is novel in the context of principal agent models and utilizes modern optimization techniques. This framework has two salient properties. First, in contrast to previous works, it does not impose any structure on the ambiguity $\mathcal{G}$ of agent technology types. Somewhat surprisingly, we can still derive useful and sometimes tight results. Second, this framework translates the study of contract optimality gap (between a particular contract family and an optimal contract) into a series of duality gaps with regard to only this said contract family, alleviating the need to compare with an optimal contract. This dramatically reduces the analytical difficulty when one wants to study the performance of affine contracts in a risk-neutral setting since expectation and affine coefficients commute. This is a key breakthrough that allows us to derive more comprehensive understanding of affine contract performance in comparison with previous literature (\eg, Tables \ref{tab:landscape-convex} and \ref{tab:landscape-non-convex}).
    \item Economic Insights of Affine Contract Performance: This paper derives succinct expressions to quantify the optimality gap of affine contracts under convex versus concave surplus functions, as well as general surplus functions based on their concave biconjugates. Two economic insights stand out. First, if agents can all learn by doing (have decreasing marginal production cost) and there exists a bottleneck agent, then affine contracts suffice. One specific instance is the ``O-ring economic production'' or orchestra performance situation, where a group of agents together produce one product, and the overall quality is determined by the bottleneck agent. Second, affine contracts tend to be better in situations where the agents have decreasing marginal cost in their production activity, as opposed to a queueing/service situation where congestion happens at high productivity or any other situation with decreasing marginal return.
    \item Complementing Existing Literature: The framework and findings complement existing literature by providing a new toolset and insights into the performance of affine contracts in an orthogonal context. This aspect is particularly relevant given the paper's discussion of works by \citet{Carroll2015, dutting2019, Yu2020, Li2022}.
    \item Insights into Technical Roots of Affine Contract (Sub)Optimality: The paper sheds light on why simple contracts often appear more optimal in robust settings compared to stochastic settings. This is attributed to the non-commutative nature of maximization and expectation operators and the fact that the sum of quasi-concave functions is not generally quasi-concave.
    \item Generalization of Decision Rules from Dynamic Robust Optimization: The study extends the analysis of decision rules, a concept from dynamic robust optimization, to the principal agent model. This approach is innovative and differs from traditional methods used in contract theory.
    \item Cross Pollination Across Disciplines: The paper's findings are relevant not only in economics but also in operations management and algorithmic game theory, indicating a wide range of potential applications and implications.
\end{enumerate}

\section{A Theoretical Framework to Bound Contract Optimality Gap under General Uncertainty Structure} \label{sec:framework}

The main problem we study comprises the principal's problem formulation (\ref{eq:DRO-principal}) and the agent's problem formulation (\ref{eq:DRO-agent}). In addition, we follow some standard assumptions in the contract theory literature.
\begin{assumption}[Compactness]\label{assumption:compact-sets-exist-optimal-solutions}
    We assume compactness for all relevant sets, thus optimal decisions and stationary points are attainable. One can safely use min and max instead of inf and sup, and also safely swap min and max when other conditions are right.
\end{assumption}

\begin{assumption}[Common Knowledge]
\label{assumption:common-knowledge}
    These information sets are common knowledge: the two players' decision problems, a plausible set of type distributions $\mathcal{G}$ called the ambiguity set, and the technology set of each technology type $\mathcal{A}_t$ for all $t$. Each $\mathcal{A}_t$ contains tuple elements like $(F, c)$ where $c$ is the cost corresponding to a specific effort level, and $F$ is the output $y$'s distribution for that effort level.
\end{assumption}

\begin{assumption}[Normalized Outside Option]\label{assumption:no-effort}
    The outside option is normalized to have 0 value for the agent. The agent can always choose to not accept the contract and choose the outside option, \ie, for any $t$, the technology set $\mathcal{A}_t$ contains the element $(\delta_0, 0)$, where $\delta_0$ is the degenerate distribution putting probability 1 on 0.
\end{assumption}

\begin{assumption}[Favorable Tie-breaking]
\label{assumption:favorable-tie-breaking}
    If the agent is indifferent among different decisions, then he will choose the one that maximizes the principal's payoff. The principal can break her ties arbitrarily. Subsequently, we slightly abuse our notations and think of ``$\arg \max$'' as a singleton instead of a set.
\end{assumption}

\begin{assumption}[Non-triviality]\label{assumption:non-triviality}
    For any $t$, there exists $(F,c) \in \mathcal{A}_t$ such that $\mathbb{E}_{F}[y] - c > 0$. This ensures that the principal can actually benefit from hiring the agent.
\end{assumption}

\begin{assumption}[One-sided Liability]\label{assumption:one-sided-liability}
The agent will never receive negative payment. In other words, we only consider contract families that contain $w(y) \geq 0$ for all $y$. In addition, we assume the mean output and the cost for each effort level are positive: $\mathbb{E}_F[y] \geq 0, c \geq 0$ for all $(F, c) \in \mathcal{A}_t$, for all $t$. 
\end{assumption}





We then define some terminologies that are useful for our analysis. 

\begin{definition}
Let $\mathcal{W}_0$ be the space of all functions consistent with our assumptions. The principal's payoff under this contract family $\mathcal{W}_0$, \ie, the objective value of \eqref{eq:DRO-principal}, is defined as the \emph{maximal principal payoff} (or simply the \emph{maximal payoff}). We call $\mathcal{W}_0$ the general contract family. Any other $\mathcal{W} \subset \mathcal{W}_0$ is called a restrictive contract family.
\end{definition}

\begin{definition}
    A contract family $\mathcal{W}$ is \emph{optimal} if the principal's payoff under this contract family is the same as the maximal principal payoff.
\end{definition}

\begin{definition}
    The \emph{optimality gap} of a contract family $\mathcal{W}$ is the difference between the maximal payoff and the principal's payoff under this contract family. Smaller optimality gaps are better.
\end{definition}

\begin{definition}
    The \emph{optimality ratio} $\leq 1$ of a contract family $\mathcal{W}$ is defined as the ratio between the principal's payoff under $\mathcal{W}$ and the maximal payoff, if the maximal payoff is strictly positive. Otherwise we say the optimality ratio is 1.
\end{definition}


It is not difficult to numerically compare the payoffs between different contract families. For example, one can directly solve the principal agent problem under each contract family case-by-case via backward induction, and then compare the principal's payoffs under the families of interest. However, such analysis usually offers limited insights if one wants to generalize them for different model settings.
Our focus in this paper is not on numerical comparisons. We instead take the approach to theoretically characterize conditions under which a restrictive family is optimal and otherwise quantify its optimality gap.

\subsection{General Results}
\label{subsec:optimality-formal}

\begin{definition}
Consider a contract family parameterized by $\theta$, $w \colon (y, \theta) \mapsto p$, where $y \in Y \subseteq \mathbb{R}$ is the output and $p \in P \subseteq \mathbb{R^+}$ is the payment. We say this contract family is \emph{\surjective{}} if $w(y,\cdot) \colon \theta \mapsto p$ is surjective with regard to $P$ for any fixed $y \in Y$.
\end{definition}
The class of \surjective{} families is fairly broad: any contract family with an offset term is a \surjective{} family. The family of output-affine contracts, $\overline{\mathcal{W}} := \{ w(\cdot, \theta, \theta_0) \, | \,  w(y, \theta, \theta_0) = \theta y + \theta_0, \theta, \theta_0 \in \mathbb{R} \}$, is \surjective{}. The set of constant payment contracts, $\dot{\mathcal{W}} := \{ w(\cdot, \theta_0) \, | \,  w(y, \theta_0) = \theta_0, \theta_0 \in \mathbb{R}^+ \}$, is \surjective{}. The family of linear contracts (affine but without an offset term) is not \surjective{} in general. 

The following theorem is the first main result of the paper. It interestingly points out that one can convert the question of optimality gap --- which is a comparison between two contract families $\mathcal{W}$ and $\mathcal{W}_0$ --- to a question about different Stackelberg games with regard to the same contract family $\mathcal{W}$.

\begin{theorem}\label{thm:main}
We define Games I, II, and III below. Let the principal's payoffs under these games be $z_{\text{I}}(\mathcal{W})$, $z_{\text{II}}(\mathcal{W})$, and $z_{\text{III}}(\mathcal{W})$. Given any \surjective{} family $\mathcal{W}$, the optimality gap of $\mathcal{W}$ is bounded above by $z_{\text{\emph{III}}}(\mathcal{W}) - z_{\text{\emph{I}}}(\mathcal{W})$. In particular, if $z_{\text{\emph{I}}}(\mathcal{W}) = z_{\text{\emph{III}}}(\mathcal{W})$, then $\mathcal{W}$ is optimal.
\end{theorem}


\paragraph{Game I:} The original game: sequence in Figure \ref{fig:sequence}. Formulations \eqref{eq:DRO-principal} and \eqref{eq:DRO-agent}.

\bigskip

\paragraph{Game II:} The principal learns the technology type before setting the contract. The game sequence is as follows.
\begin{figure}[ht]
\centering
\begin{tikzpicture}
\draw[->,ultra thick] (-6,0)--(6,0) node[right]{\small{time}};

\draw[->, thick] (-4.5,0.3)--(-4.5,0.1);
\draw[->, thick] (-1.5,0.3)--(-1.5,0.1);
\draw[->, thick] (1.5,0.3)--(1.5,0.1);
\draw[->, thick] (4.5,0.3)--(4.5,0.1);

\draw(-4.5,1.5) node[text width=2cm,align=center] {\small{technology type revealed to everyone}};
\draw(-1.5,1.5) node[text width=2cm,align=center] {\small{principal sets contract $w$}};
\draw(1.5,1.5) node[text width=2.5cm,align=center] {\small{agent accepts or declines contract; sets effort}};
\draw(4.5,1.5) node[text width=2cm,align=center] {\small{contract is executed}};
\end{tikzpicture}
\caption{Sequence of Game II.}\label{fig:sequence-II}
\end{figure}
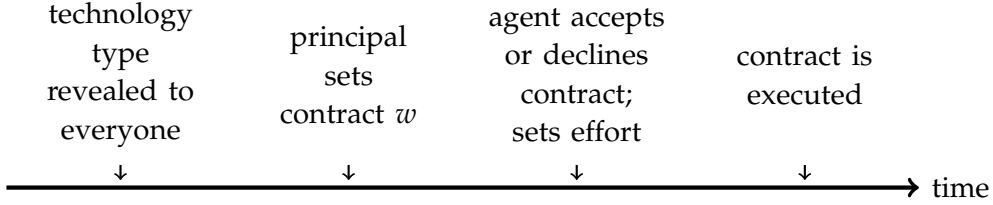
\begin{equation} \label{eq:DRO-principal-II}
    \text{Principal's Problem (II):} \quad \min_{G \in \mathcal{G}} \mathbb{E}_{t \sim G} \left[ \max_{w \in \mathcal{W}} \mathbb{E}_{y \sim F_{t,w}} [u(y) - w(y)] \right],
\end{equation}
\begin{equation} \label{eq:DRO-agent-II}
    \text{Agent's Problem (II):} \quad \text{same as \eqref{eq:DRO-agent}}.
\end{equation}


\paragraph{Game III:} The agent sets his effort level before the principal sets the payment. The game sequence is as follows. 
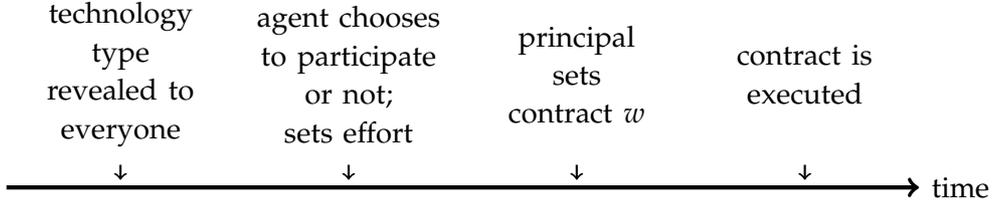
\begin{figure}[ht]
\centering
\begin{tikzpicture}
\draw[->,ultra thick] (-6,0)--(6,0) node[right]{\small{time}};

\draw[->, thick] (-4.5,0.3)--(-4.5,0.1);
\draw[->, thick] (-1.5,0.3)--(-1.5,0.1);
\draw[->, thick] (1.5,0.3)--(1.5,0.1);
\draw[->, thick] (4.5,0.3)--(4.5,0.1);

\draw(-4.5,1.5) node[text width=2cm,align=center] {\small{technology type revealed to everyone}};
\draw(-1.5,1.5) node[text width=2.5cm,align=center] {\small{agent chooses to participate or not; sets effort}};
\draw(1.5,1.5) node[text width=2cm,align=center] {\small{principal sets contract $w$}};
\draw(4.5,1.5) node[text width=2cm,align=center] {\small{contract is executed}};
\end{tikzpicture}
\caption{Sequence of Game III.}\label{fig:sequence-III}
\end{figure}

In this game, the principal chooses the payment after observing the output $y$. Define $$w_{y} :=
    \arg \max_{w \in \mathcal{W}} [u(y) - w].$$ (We treat the optimal solution set as a singleton by Assumption \ref{assumption:favorable-tie-breaking}).
    
\begin{equation} \label{eq:DRO-principal-III}
    \text{Principal's Problem (III):} \quad \min_{G \in \mathcal{G}} \mathbb{E}_{t \sim G} \left[ \mathbb{E}_{y \sim F_t} \max_{w \in \mathcal{W}} [u(y) - w] \right],
\end{equation}
\begin{equation} \label{eq:DRO-agent-III}
    \text{Agent's Problem (III):} \quad \max_{(F,c) \in \mathcal{A}_t} \mathbb{E}_{y \sim F}[w_{y}] - c,
\end{equation}
where $(F_t,c_t)$ is the optimal solution of \eqref{eq:DRO-agent-III} for type $t$ agent.


\begin{proof}[Proof of Theorem \ref{thm:main}]
    The proof directly follows from Lemma \ref{lemma:key}.
\end{proof}

\begin{lemma}\label{lemma:key}
Principal payoffs from Games I, II, III satisfy the following.
\end{lemma}
\begin{table}[h!]
    \centering
    \begin{tabular}{c c c}
         $z_{\text{I}}(\mathcal{W})$ & $\leq$ & $z_{\text{I}}(\mathcal{W}_0)$ \\
         \rotatebox[origin=c]{270}{$\leq$} & & \rotatebox[origin=c]{270}{$\leq$} \\
         $z_{\text{II}}(\mathcal{W})$ & $\leq$ & $z_{\text{II}}(\mathcal{W}_0)$ \\
         \rotatebox[origin=c]{270}{$\leq$} & & \rotatebox[origin=c]{270}{$=$} \\
         $z_{\text{III}}(\mathcal{W})$ & $=$ & $z_{\text{III}}(\mathcal{W}_0)$. \\
    \end{tabular}
\end{table}
\begin{proof}[Proof of Lemma \ref{lemma:key}]
We first prove the left-hand side column is less than or equal to the right-hand side column. We then prove the first two rows are increasing in value. We conclude the proof by showing equality between $z_{\text{III}}(\mathcal{W}_0)$ and its neighbors. 

\paragraph{Claim 1: Left-hand side column is less than or equal to the right-hand side column.} Compare the two quantities in each row in the diagram, \eg, $z_{\text{I}}(\mathcal{W})$ and $z_{\text{I}}(\mathcal{W}_0)$. The decision space for the principal's problem on the left-hand side, $\mathcal{W}$, is a subset of the decision space for the principal's problem on the right-hand side, $\mathcal{W}_0$. Thus maximization leads to a smaller objective value on the left-hand side, \ie, $z_{\text{I}}(\mathcal{W}) \leq z_{\text{I}}(\mathcal{W}_0)$. The same argument applies to all left/right pairs:
\begin{table}[ht]
    \centering
    \begin{tabular}{c c c}
         $z_{\text{I}}(\mathcal{W})$ & $\leq$ & $z_{\text{I}}(\mathcal{W}_0)$ \\
         \rotatebox[origin=c]{270}{ } & & \rotatebox[origin=c]{270}{ } \\
         $z_{\text{II}}(\mathcal{W})$ & $\leq$ & $z_{\text{II}}(\mathcal{W}_0)$ \\
         \rotatebox[origin=c]{270}{ } & & \rotatebox[origin=c]{270}{ } \\
         $z_{\text{III}}(\mathcal{W})$ & $\leq$ & $z_{\text{III}}(\mathcal{W}_0)$. \\
    \end{tabular}
\end{table}

\paragraph{Claim 2: For any fixed contract family, Game I's principal payoff is less than or equal to Game II's principal payoff.} Game I's principal payoff is less than or equal to Game II's principal payoff because the only difference between the two principal problems is the order of operations between the principal's maximization and the worst-distribution type revelation, and the rest of the bi-level problems are the same. Therefore, the principal's decision in Game II is more informed than her decision in Game I due to the observation of $G$ and $t$, and she receives a higher payoff: 
\begin{table}[ht]
    \centering
    \begin{tabular}{c c c}
         $z_{\text{I}}(\mathcal{W})$ & \, & $z_{\text{I}}(\mathcal{W}_0)$ \\
         \rotatebox[origin=c]{270}{$\leq$} & & \rotatebox[origin=c]{270}{$\leq$} \\
         $z_{\text{II}}(\mathcal{W})$ & $\,$ & $z_{\text{II}}(\mathcal{W}_0)$. \\
    \end{tabular}
\end{table}

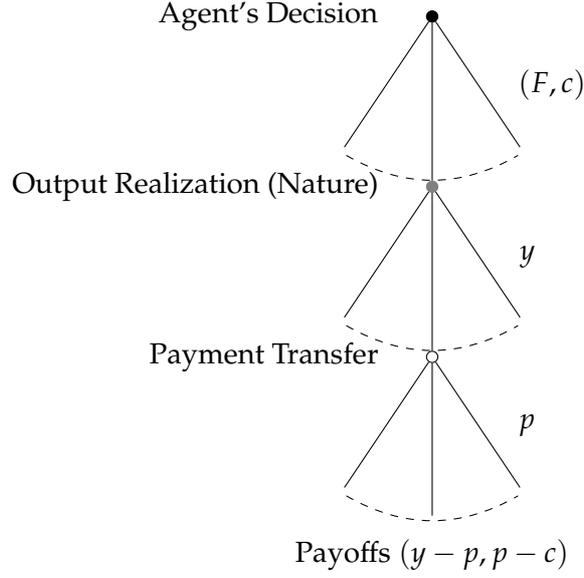
\begin{figure}[htb]
    \centering
    \tikzset{
    solid node/.style={circle,draw,inner sep=1.5,fill=black},
    hollow node/.style={circle,draw,inner sep=1.5,fill=white}
    }
    \begin{tikzpicture}[scale=1.2,font=\small]
    \tikzstyle{level 1}=[level distance=15mm,sibling distance=10mm]
    \tikzstyle{level 2}=[level distance=15mm,sibling distance=10mm]
    \tikzstyle{level 3}=[level distance=15mm,sibling distance=10mm]
    \node(0)[solid node,label=left:{Agent's Decision $\quad$}]{} 
        child{node(1)[solid node, white]{}}
        child{node(2)[solid node, gray, label=left:{Output Realization (Nature) $\quad$}, yshift=-13]{} 
            child{[black] node(4)[solid node, white, label=below:{}]{} edge from parent node[left]{}}
            child{node(5)[hollow node, label=left:{Payment Transfer $\quad$}, yshift=-13]{}
                child{node(7)[solid node, white]{}}
                child{node(9)[solid node, white, label=below:{Payoffs $(y-p, p-c)$}, yshift=-11]{}}
                child{node(8)[solid node, white, label=right:{}]{} edge from parent node[right, yshift=0, xshift = 0]{$\quad p$}}
            }
            child{[black] node(6)[solid node, white]{} edge from parent node[right]{$\quad y$}}
        edge from parent node[black, xshift=0,yshift=-42]{} 
        }
        child{node(3)[solid node, white]{} edge from parent node[right]{$\quad (F,c)$}
    };
        \draw[dashed, bend right, black](1)to(3);
        \draw[dashed, bend right, black](4)to(6);
        \draw[dashed, bend right, black](7)to(8);
    \end{tikzpicture}
    \caption{Games III can be mapped to this extensive form game (omitting the type discovery step).}
    \label{fig:extensive_II_III}
\end{figure}

\paragraph{Claim 3: $z_{\text{III}}(\mathcal{W}) = z_{\text{III}}(\mathcal{W}_0)$.} 
In Game III, the principal chooses a contract parameter after observing the actual output $y$. 
We can transform Game III to the extensive form game in Figure \ref{fig:extensive_II_III}, in which the principal's choice of contract is a choice of \emph{decision rule} mapping $y$ to $p$. 
The decision space for $p$ in the extensive form game is the range of the function $w(y,\cdot) 
\colon \theta \mapsto p$.
$P$ is the co-domain that contains all possible non-negative payment values. If we use a \surjective{} contract family, then by definition, the range $w(y,\cdot)$ equals its co-domain $P$.
This implies Game III under any \surjective{} family leads to the same principal payoff, \ie, $z_{\text{III}}(\mathcal{W}) = z_{\text{III}}(\mathcal{W}_0)$. 



\paragraph{Claim 4: $z_{\text{II}}(\mathcal{W}_0) = z_{\text{III}}(\mathcal{W}_0)$.} In Game II, the principal's decision of choosing the best decision rule from $\mathcal{W}_0$ is essentially choosing the best payment $p \in P$ as a response to every $y$ --- this is exactly in the same sense of choosing a subgame perfect solution in Game III 
with $P$ as the decision space for $p$ after observing $y$. 
More precisely, let $p^*(y)$ be the best principal response to every outcome $y$ in Game III. Then if the principal chooses this $p^*(y)$ as the contract in Game II, \eqref{eq:DRO-principal-II} and \eqref{eq:DRO-principal-III} are equal in objective value.

Combining Claims 1---4, the proof is complete.
\end{proof}




From the statements and proofs of Lemma \ref{lemma:key} and Theorem \ref{thm:main}, we know that for an arbitrary contract family $\mathcal{W}$, the optimality gap $z_{\text{III}}(\mathcal{W}) - z_{\text{I}}(\mathcal{W})$ can be broken down as 
\begin{equation}\label{eq:decompose-opt-gap}
    \text{Optimality gap of }\mathcal{W} \quad = \quad \underbrace{z_{\text{III}}(\mathcal{W}) - z_{\text{II}}(\mathcal{W})}_{\text{adjustability gap of $\mathcal{W}$}} \quad + \quad \underbrace{z_{\text{II}}(\mathcal{W}) - z_{\text{I}}(\mathcal{W})}_{\text{information rent of $\mathcal{W}$}}.
\end{equation}
We define the first part as the \emph{adjustability gap}, denoting the principal's payoff gap between two sequences: agent-then-principal versus principal-then-agent. The second part is the \emph{information rent}, capturing the (principal's) efficiency loss due to information asymmetry.

\section{Quantifying Adjustability Gap and Information Rent for Affine Contracts under Different Surplus Functions}\label{sec:theory_special_cases}

\subsection{Convex Social Surplus Functions} \label{subsection:convex_surplus}

For the next little bit we assume the output is deterministic -- agent chooses $(y,c)$ instead of $(F,c)$. We assume each $c$ maps to a unique $y$, which means there exists a production function $g(\cdot)$ such that $g(c) = y$ for each $(y,c) \in \mathcal{A}$. For simplicity, we assume $g(\cdot)$ is monotonic, but this is not crucial. We also assume the principal's utility function is linear in $y$. We shall call this model the \emph{basic model}. We will discuss relaxations at the end of the section. We define $g(c) - c$ as the \emph{social surplus function}.




We first study the adjustability gap: the principal's payoff gap between Games II and III. 
If it is clear from the context, we sometimes suppress the subscript $t$ since it is fixed.

 %

\begin{proposition}
    Under the basic model, the adjustability gap for any \surjective{} contract family $\mathcal{W}'$ is the difference between two instances of \eqref{formulation-known-type}, one with $\mathcal{W}'$ as the feasible region and one with $\mathcal{W}_0$ as the feasible region. 
    \begin{subequations}\label{formulation-known-type}
    \begin{align}
        \max_{\substack{w \in \mathcal{W} \\ (y, c) \in \mathcal{A}}} \quad & y - w(y) & \label{formulation-known-type-obj}\\
        \text{s.t.} \quad & w(y) - c \geq w(y') - c' \quad \forall (y',c') \in \mathcal{A}. & \label{formulation-known-type-IC}
    \end{align}
    \end{subequations}
    
\end{proposition}
\begin{proof}
Overall, the formulation represents the principal's problem for Game II where type is revealed before other actions. 
\eqref{formulation-known-type-obj} is the principal's objective. 
Constraint set \eqref{formulation-known-type-IC} represents the agent's optimization problem, and by the normalized outside option assumption (Assumption \ref{assumption:no-effort}), it includes both an incentive compatibility constraint and a participation constraint. By the favorable tie-breaking assumption (Assumption \ref{assumption:favorable-tie-breaking}), among the best options for the agent, he will choose one that maximizes principal payoff. Therefore, this formulation captures the bilevel optimization game. 

By the inequalities we established in Lemma \ref{lemma:key}, the following is true:
$$z_{\text{III}}(\mathcal{W}') - z_{\text{II}}(\mathcal{W}') = z_{\text{II}}(\mathcal{W}_0) - z_{\text{II}}(\mathcal{W}'),$$
where the right hand side can be compuated with two instances of \eqref{formulation-known-type} with different feasible contract regions $\mathcal{W}_0$ and $\mathcal{W}'$.
\end{proof}




Now we consider contracts that are affine and linear in $y$, which has been a topic of interest in recent literature. They are defined as follows.
\begin{subequations}
    \begin{align*}
        \overline{\mathcal{W}}_{y,0} \, & := \, \{ w(\cdot, \theta) \, | \,  w(y, \theta) = \theta_0 + \theta_1 y, \, \theta_0 \geq 0, \, \theta_1 \in [0,1] \}, \\
        \overline{\mathcal{W}}_{y} \, & := \, \{ w(\cdot, \theta) \, | \,  w(y, \theta) = \theta y, \, \theta_1 \in [0,1] \}.
    \end{align*}
\end{subequations}
Consistent with our limited liability assumption, we restrict $\theta_1$ to be from $[0,1]$ and $\theta_0 \geq 0$.

\begin{definition}[Convex Social Surplus Function]\label{def:adjustability-convex-SVF}
We say a social surplus function is convex if \Copy{convex-SVF-conditions}{$g^\prime(c) \geq 1, g^{\prime \prime}(c) \geq 0$ for $c \in [0,\bar{c}]$, and $g(0) = 0$}.
\end{definition}
Convexity conditions imply that each unit of the agent's cost can always convert into more than its value in production at all effort levels; the marginal production cost decreases as the agent puts in more effort, \eg, due to the learning-by-doing effect; and the production function is normalized at $(0,0)$. 

\begin{lemma}\label{lemma:basic_model_y_linear_family_convex_g-c}
    Suppose the social surplus function is convex. In the basic model, the adjustability gap of an affine (or linear) contract family is zero.
\end{lemma}

\begin{proof}
    First, we show that in Game III and under any \surjective{} contract family, the principal payoff in the basic model is the maximal social surplus, $\max_{c \in \mathcal{A}_c} g(c) - c$.
    
    If the production function is strictly increasing, then we can simply set the contract payment to be
    $$w(y) = g^{-1}(y),$$
    where $g^{-1}$ is the inverse of production function. The principal knows $g$ since the type is fixed and the technology set of this type is common knowledge.
    This contract pays the agent exactly the cost to generate output, so he would always receive zero. Based on the favorable tie-breaking assumption, the agent will choose an action that maximizes the principal's return, which is $\max_{c \in \mathcal{A}_c} g(c) - c$. 

    In the case where the production function is not strictly increasing, then as the principal observes output $y$, she can pay the agent the minimal cost that maps to this output level. The effective production function $\hat{g}(c)$ is strictly increasing and is without loss to the principal, since the actions that are now dominated from the agent's perspective are also dominated from the principal's perspective.
    This effectively reduces the problem to the previous case of strictly increasing production function.

    Therefore, the optimal principal payoff in Game III is the maximal point of the function $g(c) - c$. This optimal solution and corresponding optimal social surplus is the extremal point of the epograph of $g(c) - c$ in the direction of $(0,1)$ (point A in Figure \ref{fig:g-c}). We would refer to the epograph of $g(c) - c$ simply as \emph{the epograph} unless otherwise mentioned.

    In the current case of convex social surplus function, this maximal social surplus is $g(\bar{c})-\bar{c}$.
    
\begin{figure}[!htb]
    \centering
    \begin{minipage}{.5\textwidth}
        \centering
        \begin{tikzpicture}[domain=0:3.9]
        
          \draw[line width=0.25mm, ->] (-0.2,0) -- (4.2,0) node[right] {$c$};
          \draw[line width=0.25mm, ->] (0,-.8) -- (0,4.2) node[above] {$g(c)$};
        
          \draw[line width=0.25mm, dashed, color=gray] plot(\x,\x) node[right] {};
          \draw[line width=0.5mm, color=black] plot(\x,{sin(\x r) + \x * 1}) node[right] {};
        \end{tikzpicture}
        \caption{An example of (non-convex, \; non-concave) $g(c)$.}
        \label{fig:g}
    \end{minipage}%
    \begin{minipage}{0.5\textwidth}
        \centering
        \begin{tikzpicture}[domain=0:3.9]
          \draw[line width=0.25mm, ->] (-0.2,0) -- (4.2,0) node[right] {$c$};
          \draw[line width=0.25mm, ->] (0,-.8) -- (0,4.2) node[above] {$g(c)-c$};

          \draw[line width=0.5mm, color=black] plot(\x,{sin(\x r)}) node[right] {};

          \draw[line width = 0.325mm, ->] (3.14159/2,1.05)--(3.14159/2,1.4) node[right]{\tiny{$(0,1)$}};
          \filldraw[red] (3.14159/2,1) circle (1.5pt) node[below, yshift=-0.5mm]{\footnotesize{A}};

          \draw[line width = 0.325mm, ->] (3.14159/6 - 0.05/1.322, 0.5 + 0.0433/1.322)--(3.14159/6 - 0.4/1.322, 0.5 + 0.3464/1.322) node[left, xshift=-1mm, yshift=1mm]{\tiny{$(-1 + \theta_1, \theta_1)$}};
          \filldraw[blue] (3.14159/6, 0.5) circle (1.5pt) node[below, xshift=0.5mm, yshift=-0.5mm]{\footnotesize{B}};

         \draw[line width = 0.25mm, |->] (3.14159*0.75,3.14159*0.75)--(3.14159*0.75+0.28,3.14159*0.75+0.28)
         node[right]{\tiny{$(1-\theta_1, 1-\theta_1)$}};
         
        \end{tikzpicture}
        \caption{Corresponding  (non-convex, non-concave) $g(c)-c$.}
        \label{fig:g-c}
    \end{minipage}
\end{figure}
For an affine contract with parameters $(\theta_0, \theta_1)$, the agent's payoff is $\theta_1 g(c) + \theta_0 - c= \theta_1(g(c) - c) + (-1 + \theta_1)c+\theta_0$. By incentive compatibility, the agent would choose the extremal point of the epograph in the direction of $(-1 + \theta_1, \theta_1)$ (point B in Figure \ref{fig:g-c}). In the current case of convex social surplus function, since we assume $g(c)$ passes through the origin, \ie, contains the agent outside option $(0,0)$, this also satisfies individual rationality. We call this point $(c_\theta, g(c_\theta) - c_\theta)$. 

Having described and visualized the agent's decision, let us now consider the principal's decision and payoff. She wants to choose $(\theta_0, \theta_1)$ so that her payoff $(1-\theta_1)g(c_\theta)-\theta_0 = (1-\theta_1)(g(c_\theta) - c_\theta + c_\theta) -\theta_0$ is maximized. In other words, her maximization in the $g(c) - c$ versus $c$ graph is in the direction $(1-\theta_1, 1-\theta_1)$.




In the current case of convex social surplus function, if we fix $\theta_0$ at 0, then principal's optimal decision is $(\theta_0^\star, \theta_1^\star) = (0, \bar{c}/g(\bar{c}))$, which motivates the agent to pick the effort level $(\bar{c}$, leading to an output of $g(\bar{c})$ and a social surplus of $g(\bar{c}) - \bar{c})$ (Figure \ref{fig:convex-g-c}). To see this: if the principal picks some $\tilde{\theta}_1 < \bar{c}/g(\bar{c})$, then the agent will be incentivized to pick $(0,0)$ as his optimal solution, leading to the worst (zero) principal payoff; if the principal picks some $\tilde{\theta}_1 > \bar{c}/g(\bar{c})$, then the agent's decision remains at $\bar{c}$, but the principal's payoff extraction ratio, $(1 - \theta_1)$, can be increased until $\tilde{\theta}_1 = \bar{c}/g(\bar{c})$. 

Now consider $\theta_0 \neq 0$. In the current case of convex social surplus function, the agent would either pick $(0,0)$ or $(\bar{c},g(\bar{c}))$, since everything else is dominated by the line connecting these two points. The contract $(\theta_0^\star, \theta_1^\star) = (0, \bar{c}/g(\bar{c}))$ already motivates the agent to pick the high effort level and receive a total payoff of $\bar{c}/g(\bar{c}) \times g(\bar{c}) - \bar{c} = 0$, so other optimal contract with non-zero $\theta_0$ might exist but would give the same principal payoff.

In summary, the optimal decision of the principal is to set $\theta_1$ to be $\bar{c}/g(\bar{c})$, $\theta_0 = 0$, leading to the output level of $g(\bar{c})$. So the optimal principal payoff in Game II is $(1-\bar{c}/g(\bar{c}))g(\bar{c})$ = $g(\bar{c}) - \bar{c}$.

In the beginning of the proof we showed that principal payoff in Game III is equal to the maximal social surplus, $g(\bar{c}) - \bar{c}$. 

Therefore, the adjustability gap between Game II and Game III is zero. 
\end{proof}

\begin{figure}[ht]
    \centering
    \begin{minipage}{0.49\textwidth}
        \centering
        \begin{tikzpicture}[domain=0:3.75]
          \draw[line width=0.25mm, ->] (-0.2,0) -- (4,0) node[right] {$c$};
          \draw[line width=0.25mm, ->] (0,-.8) -- (0,4) node[above] {$g(c)-c$};
          \draw [dashed] (3.5, 0) -- (3.5, 4); 
          \draw [dashed] (0, 0) -- (3.5, 3.06); 
          \draw[] (3.5, -0.3) node {$\bar{c}$};
          
          \draw[line width=0.5mm, color=black] plot(\x,{\x*\x/4}) node[right] {};

          \draw[line width = 0.325mm, ->] (3.5, 3.0625)--(3.5 - 0.466*0.4, 3.0625 + 0.533 * 0.4) node[left, xshift=0mm, yshift=1mm]{\tiny{$(-1 + \theta_1^\star, \theta_1^\star)$}};
          \filldraw[blue] (3.5, 3.0625) circle (1.5pt) node[below, xshift=0.5mm, yshift=-0.5mm]{};

        \draw[line width = 0.25mm, |->] (3.14159*0.65,3.14159*0.75)--(3.14159*0.65+0.2,3.14159*0.75+0.2)
         node[left, xshift=-2mm]{\tiny{$(1-\theta^\star_1, 1-\theta^\star_1)$}};
         
        \end{tikzpicture}
        \caption{A convex social surplus function.}
        \label{fig:convex-g-c}
    \end{minipage}
    \begin{minipage}{0.49\textwidth}
        \centering
        \begin{tikzpicture}[domain=0:3.14159]
          \draw[line width=0.25mm, ->] (-0.2,0) -- (4,0) node[right] {$c$};
          \draw[line width=0.25mm, ->] (0,-.8) -- (0,4) node[above] {$g(c)-c$};

          \draw[line width=0.5mm, color=black] plot(\x,{sin(\x r)}) node[right] {};

          \draw[line width = 0.325mm, ->] (3.14159/2,1.05)--(3.14159/2,1.4) node[right]{\tiny{$(0,1)$}};
          \filldraw[red] (3.14159/2,1) circle (1.5pt) node[below, yshift=-0.5mm]{\footnotesize{A}};

          \draw[line width = 0.325mm, ->] (3.14159/6 - 0.05/1.322, 0.5 + 0.0433/1.322)--(3.14159/6 - 0.4/1.322, 0.5 + 0.3464/1.322) node[left, xshift=-1mm, yshift=1mm]{\tiny{$(-1 + \theta_1, \theta_1)$}};
          \filldraw[blue] (3.14159/6, 0.5) circle (1.5pt) node[below, xshift=0.5mm, yshift=-0.5mm]{\footnotesize{B}};

         \draw[line width = 0.25mm, |->] (3.14159*1,3.14159*0.85)--(3.14159*1+0.28,3.14159*0.85+0.28)
         node[above, xshift=0mm]{\tiny{$(1-\theta_1, 1-\theta_1)$}};
         
          \draw [dashed] (3.14*0.1, 0) -- (3.14*0.1, 4); 
          \draw[] (3.14*0.1, -0.3) node {${c}_1$};
          \draw [dashed] (3.14*0.3, 0) -- (3.14*0.3, 4); 
          \draw[] (3.14*0.3, -0.3) node {${c}_2$};
          \draw [dashed] (3.14*0.85, 0) -- (3.14*0.85, 4); 
          \draw[] (3.14*0.85, -0.3) node {${c}_3$};
         
        \end{tikzpicture}
        \caption{A concave social surplus function.}
        \label{fig:concave-g-c}
    \end{minipage}
\end{figure}



Having addressed the adjustability gap, we can now turn to the other component of the optimality gap: information rent in Eq.\ \eqref{eq:decompose-opt-gap}. Information rent comes from the difference between game sequences I and II (Fig. \ref{fig:sequence} and \ref{fig:sequence-II}). This essentially is the minimax duality gap via exchanging the maximization ($\max_{w \in \mathcal{W}}$) and worst-expectation operators ($\min_{G \in \mathcal{G}} \mathbb{E}_{t \sim G}$), \ie, formulations \eqref{eq:DRO-principal} versus \eqref{eq:DRO-principal-II}. 

A natural starting point is to look at the special case where the problem is set up as a purely robust problem: a worst-case instead of a worst-distribution consideration. In other words, the principal is interested in protecting herself against the worst type realization out of the ground set (type space) $T$. This is a special case of the distributionally robust setting because the purely robust problem is a distributionally robust problem with $\mathcal{G}$ containing $\delta_t$ for all $t \in T$, \ie, all degenerate distributions with probability 1. 

Under the purely robust setting, we show that the information rent could very possibly be zero under affine and linear contract families. This contrasts with the result that we will develop later for the Bayesian and general distributionally robust settings. 

\bigskip
\paragraph{Purely Robust Setting.} The principal's problem with the robust setting is
\begin{equation} \label{formulation:robust-principal-linear}
    \max_{\substack{\theta_0 \geq 0 \\ \theta_1 \in [0,1]}} \, \min_{t \in T} \; (1-\theta_1) g_t(c^t_\theta) - \theta_0,
\end{equation}
where $g_t$ is type $t$ agent's production function, $(c^t_{\theta}, g_t(c^t_{\theta}))$ is type $t$ agent's optimal decision under the affine contract with coefficients $\theta$.

We now establish the information rent under an affine/linear contract family for convex social surplus functions.

\begin{lemma}\label{lemma:zero-info-rent-cvx-SVF}
    Suppose the social surplus function is convex for all technology types. Under an affine (or linear) family, the information rent of the robust principal agent problem \eqref{formulation:robust-principal-linear} is zero when there is a type that is simultaneously least efficient and least productive at his highest effort level:
    $$\exists t^* \in T: t^* \in \arg \min_{t \in T} g_t(\bar{c}_t)/\bar{c}_t, \mbox{ and } t^* \in \arg \min_{t \in T} g_t(\bar{c}_t).$$
\end{lemma}
\begin{proof}
    We can write the principal's robust decision problem \eqref{formulation:robust-principal-linear} as 
    \begin{equation} \label{formulation:robust-principal-linear-simplex}
        \max_{\substack{\theta_0 \geq 0 \\ \theta_1 \in [0,1]}} \, \min_{\substack{x \in \{0,1\}^{|T|} \\ \sum_t x_t = 1}} \; \sum_{t \in T} x_t \left[(1-\theta_1) g_t(c^t_\theta) - \theta_0\right].
    \end{equation}
    This maximin problem is essentially a two-player zero-sum problem, where the two players are the principal and the nature. If we switch the sequence of play, it becomes the \emph{minimax counterpart}. The minimax counterpart of \eqref{formulation:robust-principal-linear-simplex} is 
    \begin{equation} \label{formulation:robust-principal-linear-minimax-counterpart}
        \min_{\substack{x \in \{0,1\}^{|T|} \\ \sum_t x_t = 1}} \, \max_{\substack{\theta_0 \geq 0 \\ \theta_1 \in [0,1]}} \; \sum_{t \in T} x_t \left[(1-\theta_1) g_t(c^t_\theta) - \theta_0\right].
    \end{equation}
    
    We define the minimax gap as the absolute difference in objective values between the maximin problem and its minimax counterpart (the latter is larger).

    It is clear that the information rent under an affine (or linear) family is the minimax gap of \eqref{formulation:robust-principal-linear-simplex}. 
    Thus, the proof comes down to showing the minimax gap is zero. To show the minimax gap is zero, we need to show the structure of the objective function $$\sum_{t \in T} x_t \left[(1-\theta_1) g_t(c^t_\theta) - \theta_0\right]$$ and the decision spaces are ``nice'' so a minimax equality result can hold. From the previous analysis, we know that the agent's decision $(c_\theta, g_t(c_\theta))$ in fact does not depend on the value of $\theta_0$. More precisely, formulations \eqref{formulation:robust-principal-linear} and \eqref{formulation:robust-principal-linear-simplex} can essentially be decoupled: $$\max_{\theta_0 \geq 0} -\theta_0 + \max_{\theta_1 \in [0,1]} \min_{t \in T} (1-\theta_1) g_t(c^t_{\theta_1}) = 0 + \max_{\theta_1 \in [0,1]} \min_{t \in T} (1-\theta_1) g_t(c^t_{\theta_1}).$$
    Therefore, we focus on the analysis of a linear family, and the result automatically holds for an affine family.
    
    First, we fix a type (and suppress it from the notation) and analyze the objective function with regard to the maximization player's decision $\theta_1$.
    
    Under a convex social surplus function, we previously showed (in the proof of Lemma \ref{lemma:basic_model_y_linear_family_convex_g-c}) that the principal's optimal decision is to pick $\theta_1^\star = \bar{c}/g(\bar{c})$, which motivates the agent to pick the extreme point $(\bar{c}, g(\bar{c}) - \bar{c})$ (Figure \ref{fig:convex-g-c}).
    For any $\theta_1 \in [0,\theta_1^\star)$, the agent picks $[0,0]$, which leads to a principal payoff of 0.

    For any $\theta_1 \in [\theta_1^\star,1]$, the agent picks $(\bar{c}, g(\bar{c}) - \bar{c})$, which leads to a principal payoff of $(1-\theta_1)g(\bar{c})$, which is a decreasing linear function of $\theta_1$.

    Thus, it is immediate that with a fixed type, this objective function has convex upper-level sets with regard to the argument $\theta_1$. In other words, the objective function is quasi-concave in $\theta_1$. At $\theta_1^\star$, by the favorable tie-breaking assumption that the agent always picks the decision to maximize the principal payoff if everything else is equal, this objective function is upper semicontinuous.

    For the ficticious minimization player in this game (nature), when we fix any $\theta_1$, the objective function is linear in $x$, and the decision space for $x$ is compact. 

    Overall, we have a minimax problem with compact decision spaces; the objective function is upper semicontinuous and quasi-concave in the maximization decision variable if there were only a single type, and linear in the minimization variable $x$. 

    Now write the objective function as $f(\theta, t)$.
    When there are more than one types, but there exists a ``bottleneck'' type $t^*$ that is simultaneously least efficient and least productive, 
    $t^* \in \arg \min_{t \in T} g_t(\bar{c}_t)/\bar{c}_t$, and $t^* \in \arg \min_{t \in T} g_t(\bar{c}_t)$, the preceding analysis of quasi-concavity allows us to show that at the maximizer $\theta^*_t$, the principal payoff is the least among all types, \ie, $\max_{\theta} f(\theta, t^*) \leq \min_t f(\theta^*_t, t).$

    Then the following statement is true: 
    $$\min_t \max_{\theta} f(\theta, t) \leq \max_{\theta} f(\theta, t^*) = f(\theta^*_t, t^*) \leq \min_t f(\theta^*_t, t) \leq \max_\theta \min_t f(\theta, t).$$

    Since the converse is also true by minimax inequality
    $$\min_t \max_{\theta} f(\theta, t) \geq \max_\theta \min_t f(\theta, t),$$
    we can conclude that 
    $$\min_t \max_{\theta} f(\theta, t) = \max_\theta \min_t f(\theta, t).$$
\end{proof}

\begin{theorem}
    \label{thm:cvx-SVF}
    Suppose the social surplus function is convex for all technology types. Then affine (or linear) contract family is optimal when there is a type that is simultaneously least efficient and least productive at his highest effort level:
    $$\exists t^* \in T: t^* \in \arg \min_{t \in T} g_t(\bar{c}_t)/\bar{c}_t, \mbox{ and } t^* \in \arg \min_{t \in T} g_t(\bar{c}_t).$$
\end{theorem}
\begin{proof}
The proof follows directly from Lemmas \ref{lemma:basic_model_y_linear_family_convex_g-c} and \ref{lemma:zero-info-rent-cvx-SVF}.
\end{proof}

\paragraph{Stochastic and Distributionally Robust Settings.} The stochastic and distributionally robust versions of the principal's problem under the affine contract family are
\begin{equation} \label{formulation:stochastic-principal-linear}
    \text{Stochastic:} \quad \max_{\substack{\theta_0 \geq 0 \\ \theta_1 \in [0,1]}} \, \mathbb{E}_{t \sim G_0} \; (1-\theta_1) g_t(c^t_\theta) - \theta_0,
\end{equation}
\begin{equation} \label{formulation:DR-principal-linear}
    \text{Distributionally robust:} \quad \max_{\substack{\theta_0 \geq 0 \\ \theta_1 \in [0,1]}} \, \min_{G \in \mathcal{G}} \, \mathbb{E}_{t \sim G} \; (1-\theta_1) g_t(c^t_\theta) - \theta_0.
\end{equation}

Compared with the purely robust version, these two formulations pose two technical hurdles when we try to exchange the principal's maximization operator and nature's worst-expectation operator if we want to maintain a zero minimax gap. One is the exchange of maximization with expectation. The other is the summation over different types in the objective function (in contrast, the purely robust version only picks the worst type, not multiple types). 

It is well known that maximization and expectation do not commute in general. They do only under very special conditions \citep{Yan1985}. 
This is one reason that affine and linear families are less likely to be (near) optimal for stochastic or general distributionally robust principal agent problems. 

Another interesting reason exists. If we try to generalize the result of Lemma \ref{lemma:zero-info-rent-cvx-SVF} to a distributionally robust formulation, then for any fixed nature's decision, \ie, a type distribution $G$, the objective function is a weighted sum (or integration) over different types, \ie, the sum of many $(1-\theta_1) g_t(c^t_\theta) - \theta_0$ over different $t$'s. Since the sum of quasi-concave functions is in general not quasi-concave, we can no longer prove zero information rent by the same proof logic. This negative result is true for the stochastic setting as well, since the summation/integration remains. 

\begin{observation}\label{obs:sum-quasi-concave-and-non-commutative}
Together, the fact that maximization and expectation do not commute in general, and the fact that the sum of quasi-concave functions is not quasi-concave in general shed light on the technical roots of simple families performing closer to optimality in worst-case (purely robust) rather than Bayesian (stochastic) settings.
\end{observation}

However, in some special cases, for example the one in \citet{Li2022}, production functions of different types only differ by a translation. In that case, we can show the results in \citet{Li2022} through our proof system.

\subsection{Non-Convex Social Surplus Functions}\label{subsec:non-convex-surplus}

We now characterize the adjustable gap and the information rent of affine/linear families when the social surplus function is concave for all technology types. 
When the surplus function is non-concave, we can replace $g(c)$ by its concave envelope, \ie, concave biconjugate \citep{rockafellar1997convex}, and the analysis in this section still holds if the envelope satisfies the conditions below. Concave envelope for a function $g(c)$ in $c \in [0, \bar{c}]$ is defined as the minimum of all the concave functions (defined on $[0, \bar{c}]$) that dominate $g(c)$.

\begin{definition}[Concave Social Surplus Function]\label{def:adjustability-concave-SVF}
We say the social surplus function is concave if \Copy{concave-SVF-conditions}{$g^\prime(c) \geq 1, g^{\prime \prime}(c) \leq 0$ for $c \in [0,\bar{c}]$, and $g(0) = 0$}.
\end{definition}

Intuitively, this condition means that each unit of the agent's cost can always convert into more than its value in production at all effort levels; the marginal production cost increases as the agent puts in more effort, perhaps due to resource congestion at high effort levels; and the production function is normalized at $(0,0)$. 

\begin{proposition}\label{prop:basic_model_y_linear_family_concave_g-c}
    Suppose the social surplus function is concave for a fixed type. In the basic model, the adjustability ratio of an affine (or linear) contract family is $$\frac{\max_c \; g(c)-g(c)/g^\prime(c)}{\max_c \; g(c) - c}.$$
\end{proposition}
\begin{proof}
For the case of $\bar{c} \geq \arg \max_c g(c) - c$, \eg, $\bar{c} = c_3$ in Figure \ref{fig:concave-g-c}, the problem is effectively not constrained by $[0,\bar{c}]$. The maximal social surplus is $\max_c g(c) - c$. In Game II, for any linear coefficient $\theta_1$ the principal chooses, the agent will pick a point $(c_{\theta_1},g(c_{\theta_1})-c_{\theta_1})$ such that $(-1+\theta_1, \theta_1)$ is the normal vector to the tangent direction $(1,g^\prime(c_{\theta_1})-1)$. 
This implies the relationship
$$\langle (-1+\theta_1, \theta_1), (1,g^\prime(c_{\theta_1})-1) \rangle = 0 \implies \theta_1 g^\prime(c_{\theta_1}) = 1.$$

Since the principal solves this problem 
$$\text{{maximize}}_{\theta_1} (1-\theta_1)g(c_{\theta_1}),$$
we can reformulate it via variable substitution 
$$\max_c (1-1/g^\prime(c)) g(c).$$ 
For the other two cases ($\bar{c} = c_1$ or $c_2$), it is straightforward to show that the ratio is a valid lower bound, \ie, a conservative estimation of the performance of affine and linear families.
Therefore, the adjustability ratio of an affine/linear contract family is 
$$\frac{\max_c \; g(c)-g(c)/g^\prime(c)}{\max_c \; g(c) - c}.$$
\end{proof}



As for the information rent, we can show the following.
\begin{proposition}\label{pro:principal-payoff-concave-SVF}
    Suppose the social surplus function is concave for all types. In a robust problem, for any fixed $t$, the principal's payoff under an affine contract with parameters $(\theta_0, \theta_1)$ is 
    \begin{equation}
        (1-\theta_1) \, g\left((g^\prime)^{-1}(1/\theta_1)\right) - \theta_0
    \end{equation}
    where assuming it exists, $(g^\prime)^{-1}$ is the inverse of $g'$.
\end{proposition}
The proof does not offer much insight. So we leave it out.

\begin{observation}\label{obs:concave-surplus-info-rent}
This result allows us to computationally examine each problem case by case. For example, if the social surplus function is $g(c) - c = \sin c$ for $c \in [0, \pi]$, then we can compute that the principal payoff under an affine (or linear) family is a continuous, decreasing function of $\theta_1 \in [0,1]$, satisfying the upper semicontinuity and quasi-concavity conditions. If the social surplus function is $g(c) - c = 2c^{0.5}$, then the optimal principal payoff under an affine (or linear) family does not satisfy the quasi-concavity condition. So both positive and negative cases exist for zero information rent.
\end{observation}

\subsection{Generalizations}\label{subsec:gap-model-generalization}

We made some structural assumptions in this section. Now we relax them.

\paragraph{Utility Function $u(\cdot)$.} We assumed the principal's utility function to be $u(y) = y$. Our current results directly generalize to the case of $u(y) = \alpha y + \beta$ for any $\alpha, \beta \geq 0$. For nonlinear utility functions, the overall conceptual derivation remains valid, but the precise statements of the propositions within this section would have to be modified because the principal's optimization problem is no longer linear in the $g(c) - c$ versus $c$ space.

\paragraph{Random Output and Hidden Action.} In general, our results still hold when output $y$ is random (the so-called hidden action situation). To see this, notice that under an affine or linear contract and any agent decision $(F,c) \in \mathcal{A}_t$ for some arbitrary technology type $t$, the principal's expected payoff is 
$$\mathbb{E}_{y \sim F}[y] - (\theta_0 + \theta_1 \, \mathbb{E}_{y \sim F}[y]).$$
If we revise the definition of the production function to be the maximum expected output,
\begin{equation}\label{eq:expected-g}
g(c) \, := \, \max\{\mathbb{E}_{y \sim F}[y] \, | \, (F,c) \in \mathcal{A}_t\},
\end{equation}
then the principal's payoff under an affine contract with $\theta_1 \in [0,1]$ reduces to the usual form
$$g(c) - (\theta_0 + \theta_1 \, g(c)).$$
In general, a hidden action model with random output $y$ does not change how Games I, II, and III behave under this type of affine contracts. Since our analysis of optimality (gap) mostly relies on the comparison of these three games under such affine contract families, our results with deterministic output generalize to the random output situation. 


\paragraph{General $g(c)$.} To obtain interesting closed-form results, we imposed strong conditions on $g(c) - c$ in this section: monotonicity, convexity, concavity, or worked with the concave biconjugate, and the implied continuity. But our framework in this paper does not rely on these assumptions. 

For example, even if $g(c) - c$ is discontinuous, defined on a finite domain, non-monotonic, non-convex / non-concave, we can still define the principal's optimization problem in the same way and use the epographs in the same way. For example, if the agent can only choose from a finite number of actions ($g(c)$ containing only a finite number of points in the domain) as commonly seen in the contract theory literature, then the optimization formulations remain intact, and the geometric intuition remains similar. 


\section{The Landscape of Affine Contract Performance}

\subsection{Our Results Cover Broader Situations}

We provide two comparison tables \ref{tab:landscape-convex} and \ref{tab:landscape-non-convex} that put the current paper's results and previous literature results together, to show a general landscape of the optimality (gap) of affine contracts under different situations. 

Since the setup of our theoretical framework is agnostic about the uncertainty structures, our results naturally tend to cover a broader range of cases in general, and less sharp in some specific cases. In contrast, previous results focus on specific uncertainty setups and can sometimes provide equal or sharper results for those situations, but their theoretical tools and results do not generalize. In this sense, we think the novel theoretical framework we proposed in this paper is a valuable and complementary tool for future research in contract theory, and consequently can generate complementary insights.

\begin{table}[h!]
    \footnotesize
    \centering
    \begin{tabular}{ccccc}
    \toprule
    Uncertainty     &  $t$ known & $G$ known & $G$'s mean known & $G$'s support known \\
    \midrule
    Adj Gap & 0 & 0 & 0 & 0   \\
    Reference & Lemma \ref{lemma:basic_model_y_linear_family_convex_g-c}& Lemma \ref{lemma:basic_model_y_linear_family_convex_g-c}& Lemma \ref{lemma:basic_model_y_linear_family_convex_g-c} and & Lemma \ref{lemma:basic_model_y_linear_family_convex_g-c} \\
    & & & \cite{dutting2019} \\
    \\
    Info Rent & 0 & $\neq 0$ & 0 & 0 if $\exists$ bottleneck $t$\\
    Reference & By design & Observation \ref{obs:sum-quasi-concave-and-non-commutative} & \cite{dutting2019} & Lemma \ref{lemma:zero-info-rent-cvx-SVF} and \\
    & & & & \cite{Yu2020} \\
    \bottomrule
    \end{tabular}
    \caption{A landscape for quantifying affine contract performance in different situations incorporating previous results from the literature (part 1, convex surplus function).}
    \label{tab:landscape-convex}
\end{table}

\begin{table}[h!]
    \footnotesize
    \centering
    \begin{tabular}{ccccc}
    \toprule
    Uncertainty &  $t$ known & $G$ known & $G$'s mean known & $G$'s support known \\
    \midrule
    Adj Gap & $\frac{\max_c g(c)-g(c)/g^\prime(c)}{\max_c g(c) - c}$ & $\frac{\max_c g(c)-g(c)/g^\prime(c)}{\max_c g(c) - c}$ & 0 & $\frac{\max_c g(c)-g(c)/g^\prime(c)}{\max_c g(c) - c}$   \\
    Reference & Prop. \ref{prop:basic_model_y_linear_family_concave_g-c} & Prop. \ref{prop:basic_model_y_linear_family_concave_g-c} & \cite{dutting2019} & Prop. \ref{prop:basic_model_y_linear_family_concave_g-c} \\
    \\
    Info Rent & 0 & $\neq 0$ & 0 & $\neq$ 0 \\
    Reference & By design & Observation \ref{obs:sum-quasi-concave-and-non-commutative} & \cite{dutting2019} & Observation \ref{obs:concave-surplus-info-rent}\\
    \bottomrule
    \end{tabular}
    \caption{A landscape for quantifying affine contract performance in different situations incorporating previous results from the literature (part 2, non-convex surplus function). Here $g(c)$ is the concave envelope \citep{rockafellar1997convex} of the production function.}
    \label{tab:landscape-non-convex}
\end{table}

\subsection{Reinterpretations of Some Previous Results}

Our theoretical framework can provide results in more general settings due to fewer structural assumptions, so a natural concern is that it might be less sharp in specific cases. Here we show two examples that our framework can be as sharp as some recent specialized analyses. This serves two purposes: (1) giving alternative interpretations of the previous results, and the possibility to generalize them; (2) giving confidence about our framework's sharpness in general.

\paragraph{Forest Conservation from \cite{Li2022}}

Through programs like payment for ecosystem services (PES), governments and NGOs want to incentivize forest owners to conserve forest for environmental sustainability. 
\cite{Li2022} describes a setting where the principal is a policymaker who wants to conserve forest, and the agent is a forest owner who might incur a positive opportunity cost for forest conservation. \cite{Li2022} provided an analytical framework to study the effectiveness of affine contract and its generalizations. A main finding in their work is that affine contracts are at least 50\% optimal for a quadratic cost function, and general optimality bounds can also be obtained for general convex cost functions.

We state their main setting and interpret their results in a new way. The agent has type $t \in [0,1]$, representing a nominal conservation level. The agent's action is $a$, indicating the conservation quantity choice, and $a_0$ is the total forest area. The cost of agent's action is $c(a) = \frac{h}{2}(a-t a_0)^2$ for $a \geq t a_0$, and 0 otherwise. The principal's utility is $ka$. 

In the language of our framework, we can set $y=ka$ and the principal's utility to be simply $y$. The production function is $$g(c) = k\left( \sqrt{\frac{2c}{h}} + t a_0 \right), \, g^\prime(c) = \frac{k}{\sqrt{2hc}}.$$

Now by our analysis in the concave social surplus function section, if the principal sets a linear contract with coefficient $p$, then the agent would be incentivized to use an effort level $c_p$ such that $p g'(c_p) = 1$. This implies that $c_p = \frac{1}{2} \frac{k^2}{h} p^2.$

Therefore, the principal's payoff is $$(1-p)g(c_p) = (1-p)k(p\frac{k}{h} + t a_0).$$

When $p=1/2$, it evaluates to $$\frac{1}{4} \frac{k^2}{h} + \frac{1}{2}kta_0.$$

The maximum social surplus can be calculated as 
$$\max_{c \in [0,\bar{c}]} g(c) - c \leq \max_{c \in [0,\infty]} g(c) - c = \frac{1}{2} \frac{k^2}{h} + kta_0.$$

So overall this linear contract with $p=1/2$ has an adjustability ratio greater than or equal to $1/2$.
Since this is true regardless of the type $t \in [0,1]$, the overall optimality gap of this linear contract is no worse than $1/2$ in a distributionally robust setting.



\paragraph{Salesforce Incentivization from \citet{Yu2020}}

In the salesforce incentivization setting of \citet{Yu2020} with two sales effort levels and ambiguous technology types, ``a fixed payment is sufficient to induce low effort''. The principal can add a variable payment to incentivize high effort level. Since the (expected) production function (see Eq. \eqref{eq:expected-g}) only has two points, it is essentially a linear function between low effort level and high effort level. And since the paper's setting assumes low effort level is already guaranteed, we can normalize the low effort level's cost and expected output to be $(0,0)$. This leads to two linear production functions for the two types of agents, one dominating the other, with the same function range. This domination situation is a special case of the bottlneck type condition in our Theorem \ref{thm:cvx-SVF}. Thus the adjustability gap and information rent are both zero, and an affine contract is optimal. 
The slope of the optimal contract payment coefficient can be calculated to be $$\theta_1^* = \frac{\bar{c}}{g(\bar{c})} = \frac{c(e_h) - c(e_l)}{(e_h - e_l)\left(\sum_{i=1}^n \bar{a} (y_i - y_0) - \left( \sum_{i=1}^n |\delta_i (y_i - y_0)|^q \right)^{1/q} \right)}.$$

This is the same as their main affine contract optimality result (Proposition 1) and its generalizations with $l^p$-norm ($p > 1, 1/q + 1/p = 1$) for the uncertainty set (Section 5).

\section{Conclusion}\label{sec:conclusion}

In this work, we proposed a distributionally robust formulation for the principal agent problem. It generalizes the recently popular worst-case perspective and the classical Bayesian perspective. We then developed a conceptual framework to quantify and break down the optimality gap of any parameterized, \surjective{} contract family. The framework and breakdown bring us explicit and sometimes closed-form upper bounds on the optimality gap of affine and linear families in some structured settings. We find that linear and affine families can perform well in purely robust (worst-case) settings, while the optimality of these contract families requires additional technical conditions to hold for the case of Bayesian and general \emph{distributionally} robust settings. 

\subsection{Bringing Together Literatures}

This work brings together two active streams of research. In the past decade, there have been interesting parallel developments in contract theory and robust optimization theory on the topic of robustness and simplicity. In contract theory, researchers are making breakthroughs to understand the theoretical foundations of the near-optimality of linear contracts in robust principal agent problems. In robust optimization, researchers are proving the near-optimality of linear decision rules in dynamic robust optimization problems. 

The strength of the contract theory literature is the rich modeling of information and complex incentive interactions, while the robust optimization literature usually focuses on fixed information setups and zero-sum objectives between players. The strength of the robust optimization literature is the generalized modeling of uncertainty setups and constraints, including distributional uncertainty and decision space structures. In contrast, the contract theory literature usually deals with relatively more stylized decision spaces, uncertainties, and constraints.
To bridge the two streams of literature and bring their strengths into each other, one has to generalize results from the two fields to higher levels of abstraction, in order to speak to the different types of models under a unifying lens. 

We think the most important conceptual leap that allowed us to make a connection, is the two-part realization that the optimality of (parameterized) contracts/decision rules comes down to a minimax duality issue with the appropriate reformulations; and insights developed for a minimax duality issue in a zero-sum game can be further applied in a general-sum Stackelberg setting.
We believe this is a first step in the direction of further integrating tools and insights between the robust optimization literature and the robust contract theory / robust mechanism design literature. 

\subsection{Future Research}
Many interesting questions remain. To list a few: How can we relax the two-player problem into a general agency problem where the principal may be hiring a team of agents, or the agent has downstream contracts? The former is relevant for most employer-employee settings, the latter is important for, \eg, a multi-echelon supply chain setting. How do we generalize the results to study piecewise linear contracts? These contracts also appear in practice quite often. For example, there are piecewise linear quantity-based contracts in logistics and healthcare insurance. In addition, one may be able to generalize the tools from this paper to situations with multi-dimensional payments for agents, where the multiple dimensions represent monetary compensation, healthcare benefits, family care, job location, training, job security, which are all relevant factors for a regional economy to attract talents from different socioeconomic backgrounds and grow family-sustaining jobs. 
The general framework we developed in this paper holds the potential to be a generalizable tool to study these interesting questions in future works.



%
%
%
%
%

\bibliographystyle{plainnat}
\bibliography{bibliography}

\end{document}